\documentclass[aps,prl,twocolumn,superscriptaddress,showpacs,amsmath]{revtex4}
\usepackage{graphicx}
\usepackage{bm}

\begin{document}

\title{Establishment of $\bm{M1}$ multipolarity of a 
$\bm{6.5}$~$\bm{\mu_{\text{N}}^2}$ resonance in $\bm{^{172}}$Yb at 
$\bm{E_\gamma=3.3}$~MeV}

\author{A.~Schiller}
\email{schiller@nscl.msu.edu}
\affiliation{Lawrence Livermore National Laboratory, L-414, 7000 East Avenue,
Livermore, California 94551}
\author{A.~Voinov}
\affiliation{Frank Laboratory of Neutron Physics, Joint Institute of Nuclear
Research, 141980 Dubna, Moscow region, Russia}
\author{E.~Algin}
\affiliation{Lawrence Livermore National Laboratory, L-414, 7000 East Avenue,
Livermore, California 94551}
\affiliation{North Carolina State University, Raleigh, North Carolina 27695}
\affiliation{Triangle Universities Nuclear Laboratory, Durham, North Carolina 
27708}
\affiliation{Department of Physics, Osmangazi University, Meselik, Eskisehir, 
26480 Turkey}
\author{J.A.~Becker}
\affiliation{Lawrence Livermore National Laboratory, L-414, 7000 East Avenue,
Livermore, California 94551}
\author{L.A.~Bernstein}
\affiliation{Lawrence Livermore National Laboratory, L-414, 7000 East Avenue,
Livermore, California 94551}
\author{P.E.~Garrett}
\affiliation{Lawrence Livermore National Laboratory, L-414, 7000 East Avenue,
Livermore, California 94551}
\author{M.~Guttormsen}
\affiliation{Department of Physics, University of Oslo, N-0316 Oslo, Norway}
\author{R.O.~Nelson} 
\affiliation{Los Alamos National Laboratory, MS H855, Bikini Atoll Road, Los
Alamos, New Mexico 87545}
\author{J.~Rekstad}
\affiliation{Department of Physics, University of Oslo, N-0316 Oslo, Norway}
\author{S.~Siem}
\affiliation{Department of Physics, University of Oslo, N-0316 Oslo, Norway}

\begin{abstract}
Two-step-cascade spectra in $^{172}$Yb have been measured after thermal 
neutron capture. They are compared to calculations based on experimental values
of the level density and radiative strength function (RSF) obtained from the 
$^{173}$Yb$(^3$He,$\alpha\gamma)^{172}$Yb reaction. The multipolarity of a 
$6.5(15)$~$\mu_{\text{N}}^2$ resonance at $E_\gamma=3.3(1)$~MeV in the RSF is 
determined to be $M1$ by this comparison.
\end{abstract}


\maketitle

Excited nuclei decay often by a cascade of $\gamma$ rays. While the decay 
between discrete states is determined by the details of the nuclear 
wavefunctions, unresolved transitions are best described by statistical 
concepts like a continuous radiative strength function (RSF) and level density.
The RSF (reviewed in \cite{BE73}) provides the mean value of the decay 
probability for a given $\gamma$-ray energy $E_\gamma$. For hard $\gamma$ rays,
($\sim 7$--20~MeV), the RSF is governed by the giant electric dipole resonance 
whose parameters are determined from photoabsorption \cite{DB88}. The soft tail
of the RSF has been investigated by a variety of methods, most notably by 
primary $\gamma$ rays \cite{KU95}. Recently, systematic studies of the soft RSF
have been performed at the Oslo Cyclotron Laboratory using a method based on 
sequential extraction. With this method it is possible to obtain the level
density and RSF by a deconvolution of a set of primary $\gamma$ spectra from a 
range of excitation energies \cite{SB00}. Total RSFs (summed over all 
multipolarities) of rare earth nuclei can be extracted for $B_n>E_\gamma>1$~MeV
\cite{VG01}. Their common, most striking feature is a resonance at 
$E_\gamma\sim 3$~MeV which is believed to be of dipole nature but whose 
electromagnetic character is unknown. It has been shown for all investigated 
rare earth nuclei that the total RSF is most readily decomposed into a sum of 
the Kadmenski\u{\i}-Markushev-Furman (KMF) $E1$ model \cite{KM83}, a spin-flip 
$M1$ model \cite{Ob98}, and the aforementioned soft dipole resonance 
\cite{VG01}. The knowledge of the character of this resonance is essential for 
its theoretical interpretation. Experimentally, it can be determined from a 
two-step-cascade (TSC) measurement \cite{VS03}.

The TSC method is based on the observation of decays from an initial state $i$
to a final state $f$ via one, and only one, intermediate level $m$ 
\cite{Ho58+BV91,BC95}. A convenient initial state is that formed in thermal or 
average resonance capture (ARC); the final state can be any low-lying discrete 
state. TSC spectra are determined by the branching ratios of the initial and 
intermediate states (expressed as ratios of partial to total widths $\Gamma$) 
and by the level density $\rho$ of intermediate states with spin and parity 
$J_m^\pi$
\begin{eqnarray}
\lefteqn{\hspace*{-0.4cm}I_{if}(E_1,E_2)=\sum_{XL,XL^\prime,J_m^\pi}
\frac{\Gamma_{im}^{XL}(E_1)}{\Gamma_i}\rho(E_m,J_m^\pi)
\frac{\Gamma_{mf}^{XL^\prime}(E_2)}{\Gamma_m}}\nonumber\\
&&+\sum_{XL,XL^\prime,J_{m^\prime}^\pi}
\frac{\Gamma_{im^\prime}^{XL}(E_2)}{\Gamma_i}
\rho(E_{m^\prime},J_{m^\prime}^\pi)
\frac{\Gamma_{m^\prime f}^{XL^\prime}(E_1)}{\Gamma_{m^\prime}}.
\label{eq:tsc}
\end{eqnarray}
The sums in Eq.\ (\ref{eq:tsc}) are restricted to give valid combinations of 
the level spins and parities and the transition multipolarities $XL$. They 
arise since one determines neither the ordering of the two $\gamma$ rays, nor 
the multipolarities of the transitions nor the spins and parities of the 
intermediate levels, hence one has to include all possibilities. The two 
transition energies are correlated by $E_1+E_2=E_i-E_f$, thus, TSC spectra can 
be expressed as spectra of one transition energy $E_\gamma$ only. TSC spectra 
are symmetric around $E_\gamma^{\text{sym}}=(E_i-E_f)/2$; integration over 
$E_\gamma$ yields twice the total TSC intensity $I_{if}$ if both $\gamma$ rays
are counted in the spectra. The knowledge of the parities $\pi_i$ 
\cite{footnote} and $\pi_f$ ensures that $I_{if}$ depends roughly speaking on 
the product of two RSFs around $E_\gamma^{\text{sym}}$ \cite{VS03}, i.e., 
$f_{E1}^2+f_{M1}^2$ for $\pi_i=\pi_f$ and $2\,f_{E1}\,f_{M1}$ for 
$\pi_i\neq\pi_f$. $I_{if}$ depends also on the level density. This usually 
prevents drawing firm conclusions from TSC experiments alone \cite{BC95}. A 
combined analysis of Oslo-type \textit{and} TSC experiments, however, enables 
one to establish the sum \textit{and} product, respectively, of all 
contributions to $f_{M1}$ and $f_{E1}$ at energies of the soft resonance, thus 
determining its character. For this goal, the partial widths of Eq.\ 
(\ref{eq:tsc}) are expressed via
\begin{equation}
\Gamma^{XL}_{x\rightarrow y}(E_\gamma)=f_{XL}(E_\gamma)E_\gamma^{2L+1}D_x
\label{eq:partial}
\end{equation}
in terms of RSFs and level spacings $D_x$. Eq.\ (\ref{eq:partial}) actually 
gives only the average value of the Porter-Thomas distributed partial widths 
\cite{PT56}. The total width $\Gamma$ is the sum of all partial widths. Again, 
the sum is only the sum of mean values, however, the distribution of total 
widths with many components is almost $\delta$-like \cite{PT56}. The level 
density for a given spin and parity is calculated from the total level density 
by \cite{GC65}
\begin{equation}
\rho(E_x,J_x^\pi)=\rho(E_x)\,\frac{1}{2}\,\frac{2J_x+1}{2\,\sigma^2}
\exp\left[-\frac{(J_x+1/2)^2}{2\,\sigma^2}\right],
\label{eq:cutoff}
\end{equation}
where $\sigma$ is the spin cut-off parameter, and we assume equal numbers of 
positive and negative parity levels. This assumption and Eq.\ (\ref{eq:cutoff})
have been verified from the discrete level schemes of rare earth nuclei 
\cite{GB03}. Thus, all quantities for calculating TSC spectra are based on 
experimental data.

\begin{figure}
\includegraphics[totalheight=4.3cm]{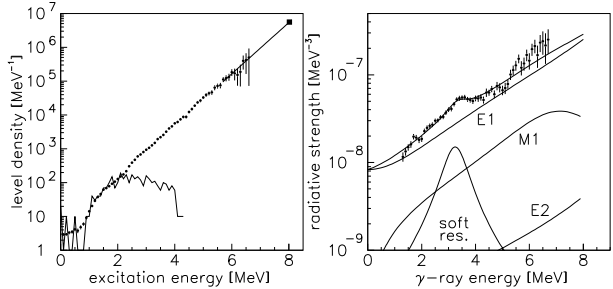}
\caption{Left panel: total level density (filled circles), constant-temperature
extrapolation (solid line), level density at $B_n$ from average neutron 
resonance spacing (filled square) \protect\cite{Ob98}, and level 
density from counting of discrete levels (jagged line) \protect\cite{FS96}. 
Right panel: total RSF (filled circles), fit to the data, and decomposition 
into RSFs of different multipolarities (solid lines). Inclusion of the soft
resonance in the fit decreases $\chi^2_{\text{red}}$ from $\sim 7.4$ to 
$\sim 1.3$. Since this value is close to unity, inclusion of additional 
non-statistical structures cannot significantly improve the fit.}
\label{fig:oslo}
\end{figure}

The combined analysis is applied to the nucleus $^{172}$Yb which has been
investigated by the $^{173}$Yb($^3$He,$\alpha\gamma)^{172}$Yb reaction in Oslo 
and by the $^{171}$Yb$(n,\gamma\gamma)^{172}$Yb reaction at the Lujan Center of
the Los Alamos Neutron Science Center (LANSCE). The Oslo data have been 
reported in \cite{SB00,VG01}. Thus, only a short summary is given. The 
experiment was performed using a 45-MeV $^3$He beam on a metallic, enriched,
self-supporting target. Ejectiles were identified and their energies measured 
using particle telescopes at $45^\circ$. In coincidence with $\alpha$ 
particles, $\gamma$ rays were detected in an array of 28 NaI detectors. From 
the reaction kinematics, $\alpha$ energy is converted into $E_x$, and $\gamma$ 
cascade spectra are constructed for a range of $E_x$ bins. The $\gamma$ spectra
are unfolded \cite{GT96} and the primary $\gamma$ spectra are extracted using a
subtraction method \cite{GR87}. The spectra are deconvoluted into a level 
density and a total RSF by applying the Brink-Axel hypothesis \cite{Br55+Ax62}.
The level density is normalized by comparison to discrete levels at low $E_x$ 
and to the average neutron resonance spacing at $B_n$ \cite{SB00}. The RSF is 
normalized using the average total width of neutron resonances, and is 
decomposed into the KMF $E1$ model, a spin-flip $M1$ model, and a soft dipole 
resonance \cite{VG01}. Here, we have improved on the normalization of the level
density and the RSF and included an isoscalar Lorentzian $E2$ model \cite{PI84}
giving
\begin{equation}
f_{\text{tot}}=K(f_{E1}+f_{M1})+E_{\gamma}^2\,f_{E2}+f_{\text{soft,}}
\label{eq:rsf}
\end{equation}
where $K$ is a scaling factor of the order of one. Since quadrupole transitions
populate levels within a broader spin interval than dipole transitions, Eq.\ 
(\ref{eq:rsf}) is of an approximative nature. Given the weakness of quadrupole 
transitions and the level of experimental uncertainties, however, this 
approximation is believed to be sufficient. The improved data, the fit to the 
total RSF, and its decomposition into different multipolarities are given in 
Fig.\ \ref{fig:oslo}. The parameters for the $E1$ RSF are taken from 
\cite{VG01}, those for the $M1$ and $E2$ RSFs from \cite{Ob98}, where we use 
the $f_{E1}/f_{M1}$ systematics at $\sim 7$~MeV giving values in agreement with
ARC work \cite{GR75}. The fit parameters are: the constant temperature of the 
KMF model $T=0.34(3)$~MeV, the normalization coefficient $K=1.7(1)$, and the 
three parameters of the soft resonance $E=3.3(1)$~MeV, $\Gamma=1.2(3)$~MeV, and
$\sigma=0.49(5)$~mb \cite{footnote2}.

\begin{figure}
\includegraphics[totalheight=8.6cm]{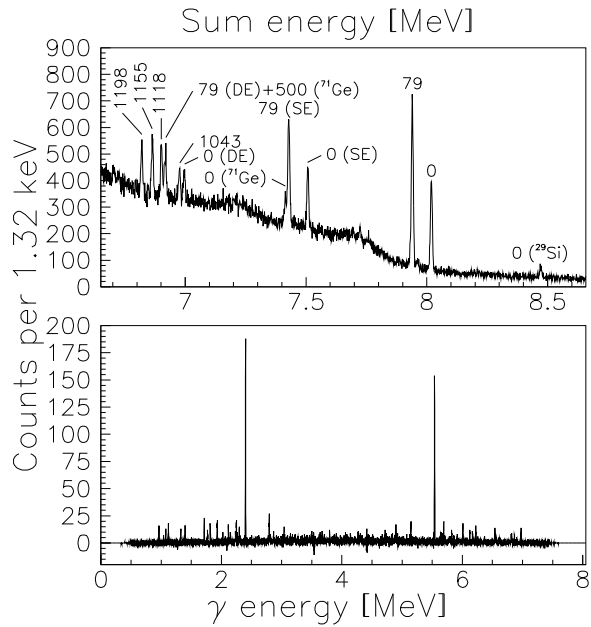}
\caption{Upper panel: energy-summed coincidence spectrum from the
$^{171}$Yb$(n,\gamma\gamma)^{172}$Yb reaction. Peaks are labeled by the energy
of the final state. Peaks denoted by $^{71}$Ge and $^{29}$Si are due to
$n$-capture in the detector and in the glass ampoule, respectively. SE and DE
stands for single and double escape peaks, respectively. Lower panel: TSC 
spectra to the $2_1^+$ state. The slight asymmetry is due to the 
energy-dependent resolution of the detectors.}
\label{fig:lansce}
\end{figure}

For the $^{171}$Yb$(n,\gamma\gamma)^{172}$Yb experiment, we used $\sim 1$~g of 
enriched, dry Yb$_2$O$_3$ powder encapsulated in a glass ampoule, mounted in an
evacuated beam tube and irradiated by collimated neutrons with a time-averaged 
flux of $\sim 4\times 10^4$ neutrons/cm$^2$s at $\sim 20$~m from the thermal 
moderator. $\gamma$ rays were detected by two 80\% and one shielded and 
segmented $\sim 200$\% clover Ge(HP) detector, placed at $\sim 12$~cm from the 
target in a geometry to minimize angular correlation effects and contributions
from higher multiplicity cascades. Single and coincident $\gamma$ rays were 
recorded simultaneously. The experiment ran for $\sim 150$~h yielding 
$\sim 10^7$ coincidences. The relative detector efficiencies from 1--9~MeV were
determined by two separate runs of $\sim 12$~h each, before and after the 
$^{171}$Yb$(n,\gamma\gamma)^{172}$Yb experiment, using the 
$^{35}$Cl$(n,\gamma)^{36}$Cl reaction and its known $\gamma$ intensities 
\cite{CB96}. Also, a standard calibrated $^{60}$Co source has been measured to 
adjust the relative curves to an absolute scale. The energy-summed coincidence 
spectrum (Fig.\ \ref{fig:lansce}, upper panel) shows distinct peaks 
corresponding to TSCs between $B_n$ and several low-lying states. The two 
strongest peaks have $\sim 4000$ counts each. TSC spectra (lower panel) were 
obtained by gating on four peaks. Relative intensities of primary versus 
secondary $\gamma$ rays were determined from singles spectra and are in 
agreement with Ref.\ \cite{GR75}. Absolute primary intensities were determined 
by using new data on absolute secondary $\gamma$-ray intensities \cite{Fi02} 
and subsequent scaling of primary intensities to these values using the 
relative intensities of \cite{GR75}. These absolute primary intensities are 
$\sim 20$\% higher than in \cite{GR75}. TSC intensities are normalized to (i) 
the absolute primary intensity and secondary branching ratio of one, strong, 
individual TSC and (ii) by effectively estimating the number of neutron 
captures during the experiment from secondary singles lines, their absolute 
intensities, and absolute detector efficiencies. Both methods give equal 
results within the error bars. 

\begin{figure}
\includegraphics[totalheight=8.6cm]{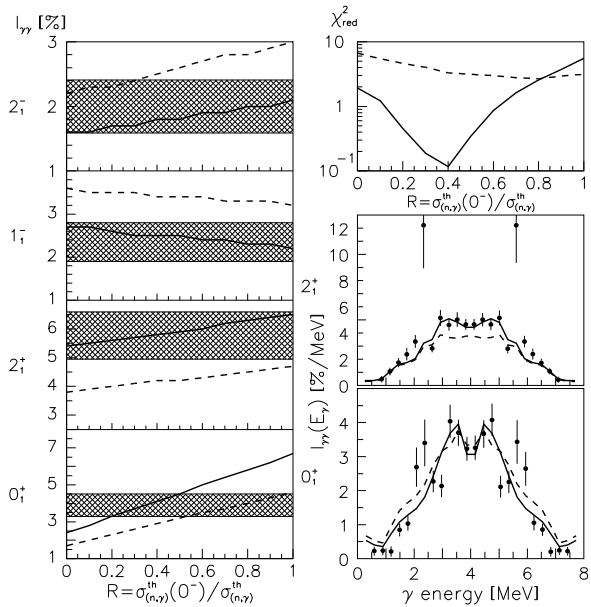}
\caption{Left: experimental values (hatched areas) for TSC intensities to final
states (from top to bottom) $2_1^-$ at 1198~keV, $1_1^-$ at 1155~keV, $2_1^+$ 
at 79~keV, and the $0_1^+$ ground state compared to calculations as function of
$R$. Included are statistical errors and systematical errors from normalization
and detection efficiency, the latter two being correlated for all final levels.
Solid and dashed lines correspond to $M1$ and $E1$ hypotheses for the soft
resonance. Right: combined $\chi^2_{\text{red}}$ for all four TSC intensities 
as function of $R$ for the $M1$ and $E1$ hypotheses (upper panel). Experimental
(filled circles) and calculated TSC spectra to the $2_1^+$ state (middle panel)
and $0_1^+$ state (lower panel) for the $M1$ hypothesis at $R=0.4$ and the $E1$
hypothesis at $R=0.8$. At $\sim 2$~MeV, Porter-Thomas fluctuations in the
experimental spectra become visible.}
\label{fig:result}
\end{figure}

TSC spectra are compared to calculations according to Eq.\ (\ref{eq:tsc}) 
assuming either electric or magnetic character for the soft resonance 
\cite{VS03}. Due to Porter-Thomas fluctuations of TSC intensities, TSC spectra 
are compressed to $\sim$300~keV energy bins and only a $\sim$2.4~MeV broad 
energy interval in the middle of the spectra is taken into account \cite{BC95} 
for comparison. Corrections due to non-isotropic angular correlations of TSCs 
have been estimated to be less than $\sim 3$\% and are thus neglected. 
Contributions to the thermal radiative neutron capture cross section 
$\sigma_{n,\gamma}^{\text{th}}$ from the two possible spins ($0^-$ and $1^-$) 
involved in neutron $s$-capture on $^{171}$Yb are uncertain. The compilation 
\cite{MD81} assumes $0^-$ for the sub-threshold resonances which contribute 
88\% to $\sigma_{n,\gamma}^{\text{th}}$. Another 4\% comes from 
$0^-$ resonances above threshold, giving in total a 92\% contribution of $0^-$ 
states. On the other hand, there is no strong evidence that all contributing 
sub-threshold resonances have $0^-$. Examination of hard primary $\gamma$-rays 
\cite{GR75,GL85} reveals many strong transitions populating $2^+$ levels, 
indicating that a sizeable portion of $\sigma_{n,\gamma}^{\text{th}}$ stems 
from $1^-$ resonances. Therefore, we performed calculations for a set of ratios
$R=\sigma_{n,\gamma}^{\text{th}}(0^-)/\sigma_{n,\gamma}^{\text{th}}$. These
calculations show, however, that only the TSC intensity to the $0_1^+$ state 
has a strong dependence on this ratio. Total experimental and calculated TSC 
intensities are shown in the left panels of Fig.\ \ref{fig:result}. The 
calculations assuming $E1$ for the soft resonance do not reproduce the 
experimental intensities consistently for any value of $R$. Good agreement is 
achieved assuming $M1$ with the additional condition of $R\sim 0.4$ for the 
$0_1^+$ final state. However, it has to be emphasized that the conclusion of an
$M1$ multipolarity for the soft resonance can be established from the TSC 
intensities to the $2_1^+$, $1_1^-$, and $2_1^-$ states independently, 
irrespective of the value of $R$. Possible systematic uncertainties in the 
absolute normalization cannot change this conclusion, since in the case of the 
final state $2_1^+$, one would need a \textit{decrease} while at the same time,
for the $1_1^-$ final state one would need an \textit{increase} in the 
experimental TSC intensities in order to accommodate the $E1$ hypothesis. The 
combined $\chi^2_{\text{red}}$ for all four TSC intensities as function of $R$ 
is also given. The $M1$ hypothesis yields the global minimum for 
$R=0.4\pm 0.25$ with $\chi^2_{\text{red}}=0.1$ whereas the minimal 
$\chi^2_{\text{red}}$ for the $E1$ hypothesis is $\sim 2.7$ for $R\sim 0.8$. 
Finally, we show the TSC spectra to two final states compared to calculations 
using the $M1$ hypothesis at $R=0.4$ and the $E1$ hypothesis at $R=0.8$. No 
further conclusions have been drawn from this comparison, however.

The integrated strength of the soft resonance is expressed as
\begin{equation}
B(M1\uparrow)=\frac{9\hbar c}{32\pi^2}
\left(\frac{\sigma\Gamma}{E}\right)_{\text{soft}}
\label{eq:bm1}
\end{equation}
giving a value of 6.5(15)~$\mu_{\text{N}}^2$ which is entirely determined from 
the Oslo-type experiment after $M1$ multipolarity has been established. This is
in agreement with the sum-rule approach for soft, orbital $M1$ strength 
\cite{LS89} but is more than twice the strength reported from nuclear resonance
fluorescence (NRF) experiments \cite{ZB90}. However, in \cite{BC95,SG00} 
several limitations in determining $B(M1\uparrow)$ using NRF are discussed, all
resulting in possible underestimation. Concerns are that (i) too few $1^+$ 
levels are observed in NRF experiments compared to level density estimates, 
(ii) the assumption in NRF experiments that the total radiative width equals 
the sum of the partial radiative widths for transitions to the ground state and
the first excited state is not fulfilled, and (iii) the excitation-energy 
coverage is insufficient. Also in \cite{BC95} a soft resonance with 
$B(M1\uparrow)\sim 7\mu_{\text{N}}^2$ is required in order to reproduce TSC 
spectra in $^{163}$Dy.

In conclusion, the soft resonance found in the RSF of $^{172}$Yb in Oslo-type 
experiments has been determined to be of $M1$ multipolarity by an auxiliary TSC
measurement. The strength of the $M1$ resonance is 
$B(M1\uparrow)=6.5(15)\ \mu_{\text{N}}^2$ which is entirely determined by the 
former experiment. This value agrees with a sum-rule approach for orbital 
strength, but is more than twice the value reported by NRF experiments. 
Assuming $M1$ multipolarity for similar soft resonances in other rare earth 
nuclei gives consistent strengths of $\sim 6\ \mu_{\text{N}}^2$ for various 
even and odd Dy, Er, and Yb nuclei and reduced strengths of 
$\sim 3\ \mu_{\text{N}}^2$ for the more spherical Sm nuclei \cite{SG02}. The 
centroids of the resonances increase weakly with mass number.

This work has benefited from the use of the Los Alamos Neutron Science Center 
at the Los Alamos National Laboratory. This facility is funded by the U.S. 
Department of Energy under Contract W-7405-ENG-36. Part of this work was 
performed under the auspices of the U.S. Department of Energy by the University
of California, Lawrence Livermore National Laboratory under Contract 
W-7405-ENG-48, and Los Alamos National Laboratory under Contract W-7405-ENG-36.
Financial support from the Norwegian Research Council (NFR) is gratefully 
acknowledged. A.V. acknowledges support from a NATO Science Fellowship under 
project number 150027/432. E.A. acknowledges support by U.S. Department of
Energy Grant No.\ DE-FG02-97-ER41042. We thank Gail F. Eaton for making the 
targets.


\begin{thebibliography}{}
\bibitem{footnote}One assumes that only neutron $s$ capture occurs.
\end{thebibliography}
\end{document}